\documentclass[iop]{emulateapj-rtx4}
\usepackage{epsfig}
\usepackage{graphicx}
\usepackage{longtable}
\usepackage{natbib}

\newcommand{\Msolar}{$M_{\odot}$}

\newcommand{\kms}{km s$^{-1}$}

\newcommand{\PRV}{$P_{RV}$}
\newcommand{\PPM}{$P_{\mu}$}
\newcommand{\vt}{$v_{t}$}
\newcommand{\logg}{log $g$}
\newcommand{\teff}{$T_\mathrm{eff}$}

\shorttitle{Barium Abundance of NGC 6819 Blue Stragglers}
\shortauthors{Milliman et al.}

\begin{document}

\title{Barium Surface Abundances of Blue Stragglers in the Open Cluster NGC 6819}

\author{Katelyn E. Milliman\altaffilmark{1}, Robert D. Mathieu\altaffilmark{1}, Simon C. Schuler\altaffilmark{2}}
\email{milliman@astro.wisc.edu}

\altaffiltext{1}{Department of Astronomy, University of Wisconsin-Madison, 475 North Charter St, Madison, WI 53706, USA}
\altaffiltext{2}{University of Tampa, 401 West Kennedy Boulevard, Tampa, FL 33606, USA}

\begin{abstract}
We present the barium surface abundance of 12 blue stragglers (BSs) and 18 main-sequence (MS) stars in the intermediate-age open cluster NGC 6819 (2.5 Gyr) based on spectra obtained from the Hydra Multi-object Spectrograph on the WIYN 3.5 m telescope. For the MS stars we find [Fe/H] = $+$0.05 $\pm$ 0.04 and [Ba/Fe] = $-$0.01 $\pm$ 0.10. The majority of the BS stars are consistent with these values. We identify five BSs with significant barium enhancement. These stars most likely formed through mass transfer from an asymptotic giant branch star that polluted the surface of the BS with the nucleosynthesis products generated during thermal pulsations. This conclusion aligns with the results from the substantial work done on the BSs in old open cluster NGC 188 that identifies mass transfer as the dominant mechanism for BS formation in that open cluster. However, four of the BSs with enhanced barium show no radial-velocity evidence for a companion. The one star that is in a binary is a double-lined system, meaning the companion is not a white dwarf and not the remnant of a prior AGB star. In this paper we attempt to develop a consistent scenario to explain the origin of these five BSs. 
\end{abstract}

\keywords{open clusters and associations: individual (NGC 6819), stars: abundances}

\section{Introduction}
Blue stragglers (BSs) are stars that do not follow the pathways laid out by standard stellar evolution. In star clusters they are typically identified as being brighter and/or bluer than the main-sequence (MS) turnoff. BSs have been identified in open and globular clusters, dwarf spheroidal galaxies, and the Galactic field. Current models invoke either mergers, collisions, or mass transfer events to explain their origins.

Extensive work has been done on the BSs in the old open cluster NGC 188 (7 Gyr). Studies show that 80\% of the BSs in NGC 188 are in binaries. Most have periods on the order of 1000 days and these long-period\footnote{In this paper long-period refers to binaries with orbital periods on the order of 1000 days or more.} systems have a secondary mass distribution peaked at 0.5 \Msolar~(\citealt{MathieuNature2009}, \citealt{GellerNature2011}). \cite{GosnellPhD} used the \textit{Hubble Space Telescope} far-ultraviolet Advanced Camera for Surveys/Solar Blind Channel to directly observe white dwarf companions of four BSs in NGC 188 and infer the presence of three more hot, young white dwarf companions (\teff~$\geqslant$ 11,000 K, age $<$ 400 Myr). All these lines of evidence indicate that mass transfer, particularly from an asymptotic giant branch (AGB) star, is the dominant formation mechanism for the BSs in NGC 188. 

Such mass transfer events can pollute the surface abundance of the BS with nucleosynthesis products from the evolved donor. Specifically, AGB mass-transfer should enhance the surface abundances of $s$-process elements, like barium, created during the thermally pulsing asymptotic giant branch (TP-AGB) phase of stellar evolution. The other formation pathways, mergers and collisions, predict no such enhancements. This makes barium an ideal marker for an AGB mass-transfer formation history of a BS.

This same scenario (i.e., surface pollution from an AGB mass transfer event) is well understood to be the formation mechanism for barium stars. Barium stars are population I red giants that have enhanced abundances of $s$-process elements (particularly barium) and carbon, compared to other giants. They make up about 1\% of all G/K-type giants. On average they have subsolar metallicities, masses between 1 \Msolar~and 2.5 \Msolar, and are in binaries (\citealt{Jorissen1998}, \citealt{Pols2003}, \citealt{Izzard2010}). 

Previous efforts to use abundance anomalies to constrain formation histories of BSs includes work on field BSs (e.g., \citealt{Sivarani2004}), globular clusters (e.g., \citealt{Ferraro2006}; \citealt{Lovisi2013}; \citealt{Lovisi2013M30}), and open clusters (e.g., \citealt{McGahee2013}; \citealt{Milliman2013}). \cite{Sivarani2004} analyzed a low-metallicity halo BS in a spectroscopic binary. They measured large overabundances of carbon, nitrogen, and oxygen as well as large enhancements in $s$-process elements, and suggest that this star accreted material from an AGB companion. \cite{Ferraro2006} discovered a subpopulation of six BSs in the globular cluster 47 Tuc that showed significant depletion of carbon and oxygen compared to the turnoff stars and the other BSs measured. \cite{Lovisi2013M30} found a similar oxygen depletion in four BSs of M30. Both ascribe these depletions to the products of CNO burning being on the surface of the BS and conclude these BSs to have a mass transfer origin with a red giant donor. 

For this paper, we focus on identifying an AGB mass transfer origin for BSs by measuring the barium abundance of BS and MS stars in the rich intermediate-age open cluster NGC 6819 (2.5 Gyr; $\alpha$= 19$^\mathrm{h}$41$^\mathrm{m}$17$^\mathrm{s}$.5 (J2000), $\delta$= $+$40$^{\circ}$11$'$47$''$). As one of four open clusters located within the original field of view of \textit{Kepler}, NGC 6819 has experienced an increase in interest in recent years. For example, the WIYN Open Cluster Study\footnote{This is WIYN Open Cluster Study paper LXVIII} (WOCS; \citealt{Mathieu2000}) has published wide-field $VI$ photometry (\citealt{Yang2013}), proper motions (PM, \citealt{Platais2013}), radial velocities (RV, \citealt{Milliman2014}), and iron abundance measurements (\citealt{LeeBrown2015}) for this cluster. Other work has uncovered several peculiar objects in NGC 6819 including a candidate quiescent low-mass X-ray binary (\citealt{Gosnell2012}), possible evolved descendants of past BSs (\citealt{RV1998}, \citealt{Corsaro2012}), and a puzzling lithium-rich red giant (\citealt{AnthonyTwarog2013}, \citealt{Carlberg2015}). The extensive information available for this cluster combined with its rich BS population made NGC 6819 an ideal target for our BS abundance study.

\section{Target Selection}
\begin{figure}[htbp]
\begin{center} 
\includegraphics[width=0.95\linewidth]{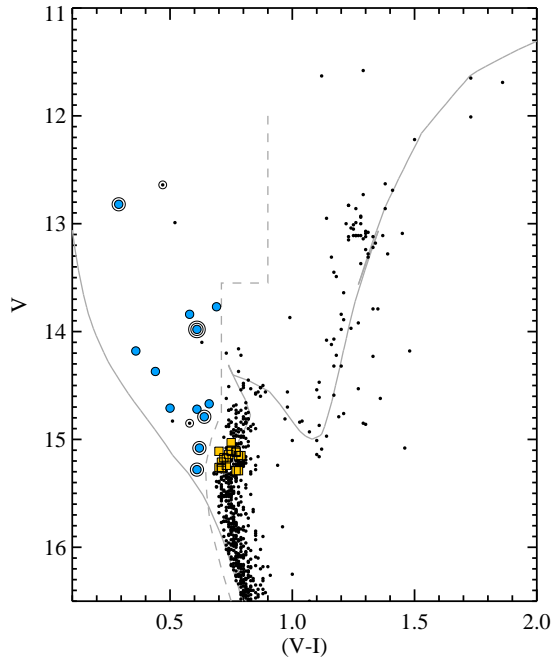}
\caption{CMD of all radial-velocity and proper-motion members of NGC 6819. To guide the eye a PARSEC zero-age main-sequence and a 2.5 Gyr isochrone are shown as the gray solid lines. Stars bluer than the dashed line are considered blue stragglers. Binary blue stragglers are marked by a circle and the one double-lined system is marked by two circles. Stars with measured abundance values are marked with large blue circles for the blue stragglers and large orange squares for our main-sequence control sample.}
\label{fig:cmd}
\end{center}
\end{figure}

We selected targets for our abundance study that are identified as members of NGC 6819 based on PM (\citealt{Platais2013}) and RV (\citealt{Milliman2014}) information. \cite{Milliman2014} identified 17 BSs in NGC 6819 that are three-dimensional kinematic members, but that fall outside of the magnitude and color limits for normal stars and binaries from the cluster turnoff region or blue hook. We were able to obtain abundance measurements for 12 of these 17 BSs. Rapid rotation or fiber separation requirements prevented us from getting measurements for the complete sample of BSs. To compare our BS sample to the standard abundance value in NGC 6819 we also obtained measurements for 18 MS stars that are single cluster members with narrow spectral lines. Figure \ref{fig:cmd} shows the distribution of these targets in the color-magnitude diagram (CMD) for NGC 6819 using $VI$ photometry from \cite{Yang2013}.

%Figure 1 fig.cmd.eps

Table~\ref{tab:specs} lists the photometry, membership probabilities, $e/i$, and membership classification for the stars in our sample. For a full explanation of $e/i$ please see \cite{Geller2008}, but briefly it is the ratio of the RV standard deviation (the external error, $e$) to the measurement precision (the internal error, $i$). The WOCS RV surveys use an $e/i$ threshold of 4 to distinguish between velocity-variable stars and single members (SMs). The membership classification category is explained in detail in \cite{Milliman2014} and the distinctions are: cluster members have \PRV~$\geqslant$ 50\% and \PPM~$\geqslant$ 4\%; SMs have $e/i$ $<$ 4.0; binary likely members (BLMs) have an $e/i$ $\geqslant$ 4.0, but do not have a completed orbital solution; binary members (BMs) have an $e/i$ $\geqslant$ 4.0 and a complete orbital solution. \cite{Milliman2014} find $\sim$40\% (7/17) of the BSs in NGC 6819 to be either BLMs or BMs. We are able to measure surface abundance values for five of these velocity-variable BSs. The two remaining binary BSs are too rapidly rotating ($v$sin$i$ $\cong$ 60 \kms~and 70 \kms) for us to get accurate abundance measurements.  

\begin{deluxetable*}{lcccccccccc}
\tablewidth{\textwidth}
\centering
\tabletypesize{\scriptsize}
\tablecaption{Stellar Information and Parameters\label{tab:specs}}
\tablehead{ \colhead{WOCS ID} & \colhead{} & \colhead{$V$} & \colhead{$V-I$} &\colhead{\PRV} & \colhead{\PPM} & \colhead{$e$/$i$} & \colhead{Class} & \colhead{$T_{eff}$ (K)} & \colhead{log $g$} & \colhead{$v_{t}$ (\kms)}}
\startdata
 1010 &      BS & 12.82 &  0.29 & 86 &      87 &   8.66 &  BM & 8524 & 3.75 &  1.0 - 4.0 \\
 4004\tablenotemark{a} &      BS & 13.98 &  0.61 & 94 &      99 &  17.45 &  BM & 6900 & 3.95 & 2.0 - 2.5 \\
  &  &  &   &  &  &  & & 6650 & 4.0 & 2.0 \\
 5006 &      BS & 13.84 &  0.58 & 94 &      99 &   1.75 &  SM & 6970 & 3.75 &  1.7 - 2.0 \\
 8021 &      BS & 14.18 &  0.36 & 93 &      15 &   1.29 &  SM & 8006 & 4.13 &  2.0 - 3.0 \\
 9003 &      BS & 13.77 &  0.69 & 94 &      99 &   0.93 &  SM & 6591 & 3.60 &  2.0 - 2.5 \\
10006 &      BS & 14.37 &  0.44 & 79 &      97 &   1.98 &  SM & 7419 & 4.09 &  2.0 - 3.0 \\
14015 &      BS & 14.71 &  0.50 & 92 &      97 &   2.28 &  SM & 7480 & 4.14 &  2.0 - 3.0 \\
16009 &      BS & 14.79 &  0.64 & 93 &      99 &   5.72 & BLM & 6679 & 3.97 &  2.0 - 2.5 \\
16021 &      BS & 14.72 &  0.61 & 84 &       6 &   1.94 &  SM & 6820 & 4.00 &  2.0 - 2.5 \\
21022 & \nodata & 15.12 &  0.77 & 86 &      47 &   1.89 &  SM & 6282 & 3.96 &  2.0 - 2.5 \\
22018 & \nodata & 15.03 &  0.75 & 86 &      93 &   0.43 &  SM & 6352 & 3.90 &  2.0 - 2.5 \\
22022 & \nodata & 15.14 &  0.74 & 93 &      53 &   0.40 &  SM & 6390 & 3.96 &  2.0 - 2.5 \\
23017 & \nodata & 15.11 &  0.75 & 94 &      99 &   1.19 &  SM & 6353 & 3.96 &  2.0 - 2.5 \\
24027 & \nodata & 15.15 &  0.79 & 94 &      58 &   0.88 &  SM & 6248 & 3.96 &  2.0 - 2.5 \\
25027 & \nodata & 15.26 &  0.70 & 93 &      61 &   1.09 &  SM & 6508 & 4.00 &  2.0 - 2.5 \\
26017 & \nodata & 15.17 &  0.72 & 93 &      99 &   3.21 &  SM & 6467 & 3.96 &  2.0 - 2.5 \\
26019 & \nodata & 15.21 &  0.71 & 94 &      75 &   2.83 &  SM & 6549 & 3.98 &  2.0 - 2.5 \\
27019 & \nodata & 15.24 &  0.73 & 94 &      98 &   0.67 &  SM & 6352 & 4.00 &  2.0 - 2.5 \\
27021 & \nodata & 15.18 &  0.73 & 84 &      57 &   1.24 &  SM & 6389 & 3.96 &  2.0 - 2.5 \\
28017 & \nodata & 15.27 &  0.71 & 93 &      99 &   1.84 &  SM & 6389 & 4.00 &  2.0 - 2.5 \\
30027 & \nodata & 15.29 &  0.78 & 57 &      56 &   0.32 &  SM & 6317 & 4.02 &  2.0 - 2.5 \\
31028 & \nodata & 15.10 &  0.75 & 94 &      60 &   1.57 &  SM & 6248 & 3.96 &  2.0 - 2.5 \\
32023 &      BS & 15.28 &  0.61 & 94 &      60 &   4.73 & BLM & 6869 & 4.19 &  2.0 - 2.5 \\
32026 & \nodata & 15.11 &  0.70 & 93 &      59 &   0.73 &  SM & 6428 & 3.96 &  2.0 - 2.5 \\
33038 & \nodata & 15.15 &  0.78 & 92 &      36 &   1.06 &  SM & 6282 & 3.96 &  2.0 - 2.5 \\
41039 & \nodata & 15.11 &  0.77 & 94 &      58 &   0.37 &  SM & 6352 & 3.96 &  2.0 - 2.5 \\
47030 &      BS & 15.08 &  0.62 & 92 &      21 &   8.01 & BLM & 6820 & 4.11 &  2.0 - 2.5 \\
51059 &      BS & 14.67 &  0.66 & 53 & \nodata &   1.19 &  SM & 6635 & 3.95 &  2.0 - 2.5 \\
62034 & \nodata & 15.26 &  0.71 & 93 &      47 &   1.53 &  SM & 6316 & 4.00 &  2.0 - 2.5 \\
63044 & \nodata & 15.29 &  0.77 & 85 &      47 &   0.27 &  SM & 6123 & 4.02 &  2.0 - 2.5 
\enddata
\tablenotetext{a}{WOCS ID 4004 is a double-lined system. The spectral parameters for the primary are listed in the first entry and the secondary information is listed second.}
\end{deluxetable*}

\section{Observations and Data Reduction}
We obtained spectra of NGC 6819 in 2014 July with the Hydra Multi-object Spectrograph (\citealt{Barden1994}) on the WIYN 3.5 m telescope. Our observations included two different spectrograph setups covering two different wavelength regions centered at 5835 \AA~and 6646 \AA. The first setup has a wavelength range from 5700 to 5950 \AA~with a resolution of $R$ $\sim$ 19,000 and total exposure time of 5.5 hr. For these spectra we achieved a signal-to-noise ratio of 95 per resolution element for the faintest MS star ($V$ = 15.29) and over 300 for the brightest BS ($V$ = 12.82). The second spectral range extends from 6440 to 6840 \AA~with a resolution of $R$ $\sim$ 13,000 and total exposure time of 8.7 hr. We achieved a signal-to-noise ratio in this spectral range of 150 per resolution element for the faintest MS star and over 415 for the brightest BS. The results of this paper focus on the measured barium abundance from the \ion{Ba}{2} feature at 5853.7 \AA~and our iron abundance was based on \ion{Fe}{1} lines from both spectral windows.   

The data were reduced using standard IRAF\footnote{IRAF is distributed by the National Optical Astronomy Observatories, which are operated by the Association of Universities for Research in Astronomy, Inc., under cooperative agreement with the National Science Foundation.} routines and following the basic procedure outlined in \cite{Steinhauer2003}. In brief, the individual frames were trimmed and overscan subtracted before a master bias, made from combining a series of bias frames, was subtracted. Cosmic rays were removed from the object frames using the routine L.A. Cosmic (\citealt{Dokkum2001}). The spectra were then dispersion corrected, and the extracted spectra were flat-fielded, throughput corrected, and sky subtracted. Sky levels were determined from 27 sky fibers placed throughout the field of view that were averaged together to create a single sky spectrum. We corrected for Doppler shift using RVs derived from the centroid of a one-dimensional cross-correlation function with the $fxcor$ task and an observed day-time sky spectrum. Individual exposures were added together to create the final spectra used in this analysis.

\section{Atmospheric Parameters}
\label{sec:params}
As is typical for abundance studies at these spectral resolutions (e.g.~\citealt{AnthonyTwarog2010}), we obtain atmospheric parameters based on photometric relationships and isochrones that we detail below.

We determined the stellar parameters listed in Table~\ref{tab:specs} after adopting the reddening and distance modulus from \cite{Yang2013}; $E(B-V)$ = 0.14, $E(V-I)$ = 0.20, and $(m-M)_V$ = 12.36. Effective temperatures were based on the $BV$ color from \cite{Hole2009} and the following relationship from \cite{BLG1998},
\begin{eqnarray} 
T_{\mathrm{eff}} &=& \big[9134 - 8600(B-V) + 5398(B-V)^2\big] \nonumber \\
        & & \Bigg[1+\frac{0.40+1.30\mathrm{[Fe/H]}}{100}\Bigg].
\label{eq:Tblg}
\end{eqnarray}
We used the \cite{Brag2001} value of [Fe/H] = $+$0.09 for the metallicity. Adopting [Fe/H] = $-$0.02 from \cite{LeeBrown2015} changes our temperatures by 1.2\% or approximately 75 K for our MS stars. For the three stars without $BV$ information (WOCS 33038, WOCS 41039, and WOCS 51059), temperature was based on stars nearby in the $VI$ CMD that did have $BV$ photometry. 

We adopted \logg~values from a 2.5 Gyr PARSEC (\citealt{PARSEC2012}) isochrone for the MS stars and for the BSs we used PARSEC evolutionary tracks that passed through the $L$-\teff~positions of these stars. We recognize this method for approximating BS gravity is simplistic based on uncertainties in BS masses (\citealt{MathieuGeller2015}), but we take this into account by adjusting the \logg~values by $\pm$0.2 dex in our analysis. We also note that [Fe/H] shows little dependence on \logg~and that microturbulence, \vt, has a much stronger effect on [Ba/Fe] than \logg~(see Table \ref{tab:shifts}).

For the MS stars and fainter BSs we varied the microturbulence velocity from 2.0 \kms~to 2.5 \kms~based on the \vt~results from \cite{Edvardsson1993} and \cite{Ramirez2013}. Based on the \cite{Takeda2008} study of A-type stars we chose a greater microturbulence range for the BSs in our sample with temperatures hotter than 7400 K. For WOCS 8021, WOCS 10006, and WOCS 14015 we varied \vt~from 2.0 \kms~to 3.0 \kms. For the hottest star in our sample, WOCS 1010, we varied the microturbulence from 1.0 \kms~to 4.0 \kms. BS WOCS 5006 has enough good quality iron lines that we could determine the range for \vt~to be 1.7 \kms~to 2.0 \kms. The values for microturbulence that we used for each star are listed in Table~\ref{tab:specs}.

WOCS 4004 is a special case in that it is a double-lined spectroscopic binary (SB2). The spectra and photometry have significant light contribution from both the primary and secondary star. For the atmospheric parameters we adopted the primary and secondary temperature values of \cite{Milliman2014} that were determined from cross correlating the spectra of WOCS 4004 against a grid of synthetic spectra. Surface gravities were based on a comparison between the temperature for each component, PARSEC evolutionary tracks, and the composite color and magnitude of WOCS 4004. We varied the microturbulence velocity from 2.0 \kms~to 2.5 \kms. 

In order to establish that our abundance results are significant and not highly dependent on our adopted parameters we vary each quantity by an amount larger than the expected error. We shift \teff~by $\pm$200 K, \logg~by $\pm$0.2 dex, and \vt~by +0.5 \kms~or more. The effects of each of these adjustments are discussed in more detail in Section~\ref{sec:results} and listed in Table~\ref{tab:shifts}.

\section{Abundance Analysis}
Our analysis was done using the 2014 July version of the LTE analysis code MOOG\footnote{http://www.as.utexas.edu/$\sim$chris/moog.html} (\citealt{Sneden1973}) incorporating the ATLAS9 model atmospheres of \cite{ATLAS9}. Our line list was compiled based on atomic information from the Vienna Atomic Line Database, (\citealt{VALD2000}). We adjusted the log \textit{gf} values slightly for two iron lines in order to better match our observed solar spectrum\footnote{For our solar spectrum we combined together day-time sky spectra taken with the same fiber configuration and reduced with the same methods as our science spectra.} to expected abundance values. We also included hyperfine structure for the \ion{Ba}{2} line at 5853.7 \AA~as this better matched our day-time sky spectra with the expected solar abundance value. 

We measured the equivalent widths of our iron lines using the spectrum analysis package SPECTRE\footnote{http://www.as.utexas.edu/$\sim$chris/spectre.html}. We used these equivalent widths and the MOOG \textit{abfind} routine to derive the abundance of each iron line. The [Fe/H]\footnote{We use standard bracket notation to denote stellar abundances relative to solar values, e.g., [m/H] = log[$N$(m)/$N$(H)]$_{\star}$ $-$ log[$N$(m)/$N$(H)]$_{\odot}$ on a scale where log $N$(H) = 12.0.} abundance for each star was then determined by averaging the line-by-line differential abundances, i.e., the average of all log $N_{\star}$(\ion{Fe}{1})$_{line~i}$ $-$ log $N_{\odot}$(\ion{Fe}{1})$_{line~i}$ for each star. The number of iron lines measured for each star is listed in Table~\ref{tab:abund.info}. To determine the barium abundance we fit synthetic spectra to our observed spectra using the MOOG $synth$ routine, focusing on the \ion{Ba}{2} line at 5853.7 \AA. Based on a solar model with atmospheric parameters \teff~= 5777 K, \logg~= 4.44, \vt~= 1.38 \kms, and [Fe/H] = 0, we obtained log $N_{\odot}$(\ion{Fe}{1}) = 7.602 and log $N_{\odot}$(Ba) = 2.18. We determined the barium abundance relative to iron for each star using: [Ba/Fe]= (log $N_{\star}$(Ba) $-$ log $N_{\odot}$(Ba))$-$[Fe/H]$_{\star}$.

\cite{Korotin2011} investigated NLTE effects on the [Ba/Fe] abundance of dwarf stars with similar temperatures, gravities, and [Fe/H] to the stars in our sample and find a systematic increase of only $\sim$0.05 dex. Based on this result we do not consider NLTE effects in this paper.

\subsection{The Double-lined System WOCS 4004}
\label{sec:4004}
WOCS 4004 is an SB2 with a period of 297 days, an eccentricity of 0.21, and a mass ratio, $m_{2}$/$m_{1}$, of 0.73 (\citealt{Milliman2014}). We used a two-dimensional correlation technique, TODCOR, developed by \cite{Zucker1994} to determine the RVs of both the primary and secondary component. The two stars were separated by 30 \kms~at the time of the observation. We converted the spectra to a rest wavelength based on the RV of the primary star. As mentioned in Section~\ref{sec:params}, the individual effective temperatures of the primary and secondary stars were derived from cross-correlation with a grid of synthetic spectra (\citealt{Milliman2014}). The intensity ratio, $F_{2}$/$F_{1}$, of 0.35 $\pm$ 0.04 was derived using a similar technique from spectra centered at 5187 \AA~by Dr. G. Torres (2013, private communication). 

\begin{figure}[htbp]
\begin{center} 
\includegraphics[width=0.95\linewidth]{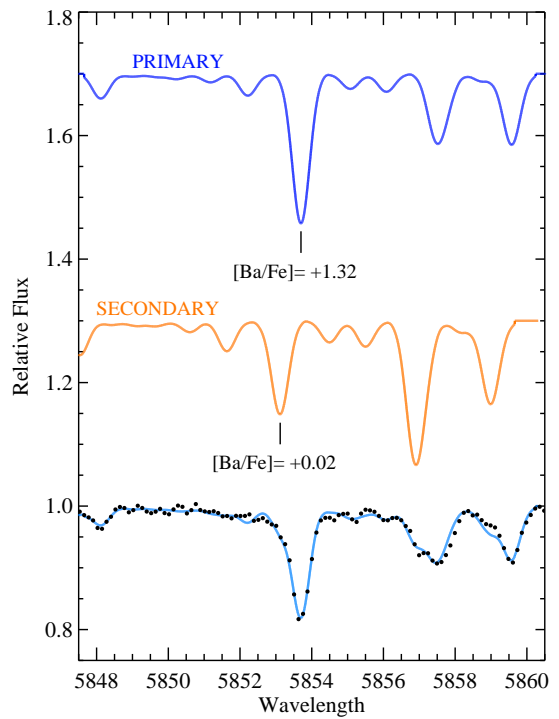}
\caption{For the double-lined binary blue straggler, WOCS 4004, we are able to use the primary and secondary temperatures derived by \cite{Milliman2014}, the luminosity ratio provided by Dr. G. Torres (2013, private communication), and TODCOR RVs to fit the observed spectrum (black dots) and derive the barium abundance for the individual stars. The best fit synthetic spectra (light blue) combines a hot primary (\teff~= 6900 K, dark blue) with significant barium enhancement and a cooler secondary (\teff~= 6650 K, orange) with a barium abundance consistent with the expected cluster value.}
\label{fig:4004}
\end{center}
\end{figure}

Because the absorption lines in the observed spectra have contributions from the primary and secondary stars determining [Fe/H] abundance from equivalent width measurements is problematic. Therefore we determined both [Fe/H] and [Ba/Fe] by fitting synthetic spectra to our observed spectra using the MOOG $synth$ routine. For these fits we assumed the [Fe/H] values for the primary and secondary to be equal. To obtain our abundance measurements we adjusted the iron and barium values of the synthetic spectra representing the primary and secondary stars until the combined spectrum best matched the observed spectrum as shown in Figure~\ref{fig:4004}. 

%Figure 2; 4004.spec.ps

As described in Section~\ref{sec:results} we repeated this fitting procedure with a variety of temperature, gravity, and microturbulence values for the primary's synthetic spectrum. In order to limit the number of free parameters and because the secondary star showed no unusual abundance features the atmospheric values for the secondary star were not adjusted. For each of the six best-fit synthetic spectra the iron and barium abundances of the secondary were near solar, but the barium abundance of the primary was significantly above solar (see Table~\ref{tab:abund.info}).

\section{Abundance Results}
\label{sec:results}
For our adopted atmospheric parameters, assuming a \vt~= 2.0 \kms, the MS stars in our sample have an average [Fe/H] = $+$0.05 $\pm$ 0.04 (the standard deviation of the distribution, s.d.). This value is consistent with the abundance \cite{Brag2001} found from red clump stars, [Fe/H] = $+$0.09 $\pm$ 0.03, and close to the value \cite{LeeBrown2015} found from MS and turn-off stars, [Fe/H] = $-$0.02 $\pm$ 0.02. We find the average barium value for the MS sample is roughly solar, [Ba/Fe] = $-$0.01 $\pm$ 0.10 (s.d.). For the entire sample of BSs the average [Fe/H] = $+$0.06 $\pm$ 0.09 (s.d.). This is consistent with the above values, but with larger scatter perhaps caused by the higher temperature range of the BSs or because the atmospheric parameters are based on MS relationships and might be farther from the true BS values.

To ensure that our abundance results are robust and are not sensitive to our adopted atmospheric parameters we varied each parameter by an amount that exceeded the expected error. Specifically, for each star we shifted \teff~by $\pm$200 K, log~$g$ by $\pm$0.2 dex, and varied the \vt~values as detailed in Section~\ref{sec:params}. Table~\ref{tab:shifts} outlines the average change in [Fe/H] and [Ba/Fe] from shifting the temperature, gravity, and microturbulence values for the 18 MS stars in our sample. Temperature has the greatest impact on our [Fe/H] results, changing the abundance by a little over 0.1 dex. Our [Ba/Fe] results were most effected by the change in microturbulence, with the 0.5 \kms~increase resulting in 0.13 lower [Ba/Fe] values.

\begin{deluxetable*}{lccccc}
\tablewidth{\textwidth}
\centering
\tabletypesize{\footnotesize}
\tablecaption{Abundance Shifts due to Atmospheric Parameters\label{tab:shifts}}
\tablehead{ \colhead{} & \colhead{\teff} & \colhead{\teff} & \colhead{\logg}    & \colhead{\logg}    & \colhead{\vt} \\
            \colhead{} & \colhead{($+$200 K)}  & \colhead{($-$200 K)}  & \colhead{($+$0.2 dex)} & \colhead{($-$0.2 dex)} & \colhead{($+$ 0.5 \kms)}}
\startdata
$[$Fe/H$]$  & $+$0.11 & $-$0.11 & $-$0.01 &  $+$0.00 & $-$0.05 \\ 
$[$Ba/Fe$]$ & $-$0.02 & $+$0.04 & $+$0.08 &  $-$0.07 & $-$0.13
\enddata
\end{deluxetable*}

\begin{figure}[htbp]
\begin{center} 
\includegraphics[width=\linewidth]{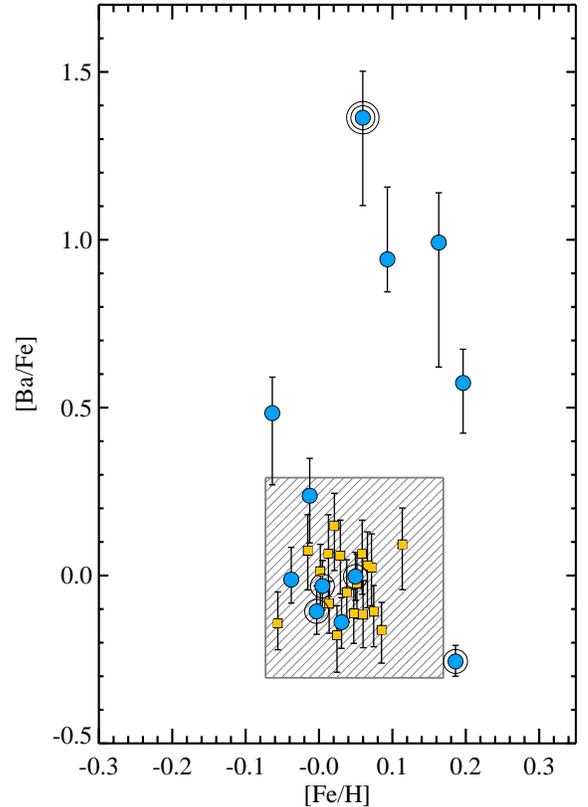}
\caption{[Ba/Fe] vs. [Fe/H] for the stars in our sample. Blue stragglers are marked by blue circles and main-sequence stars are marked by orange squares. Binarity is marked by open circles and the primary of the SB2 WOCS 4004 is marked by two circles. The gray hatched box marks the 3$\sigma$ variation region of the MS values. The error bars mark the amount of variation in the [Ba/Fe] values from adjusting the temperature, \logg, and \vt~values for all of the stars.}
\label{fig:abund}
\end{center}
\end{figure}

With these changes in atmospheric parameters we measured six [Fe/H] and [Ba/Fe] values for each star. The average of these six values is presented in Table~\ref{tab:abund.info}, along with the minimum and maximum [Fe/H] and [Ba/Fe] values for each star. In Figure~\ref{fig:abund} we plot the average [Fe/H] versus average [Ba/Fe] for the MS and BS stars and use error bars to indicate the complete range in measured [Ba/Fe]. We also overplot the region corresponding to the 3$\sigma$ variation of the MS abundance values.

\begin{figure*}[htbp]
\begin{center} 
\includegraphics[width=0.8\linewidth]{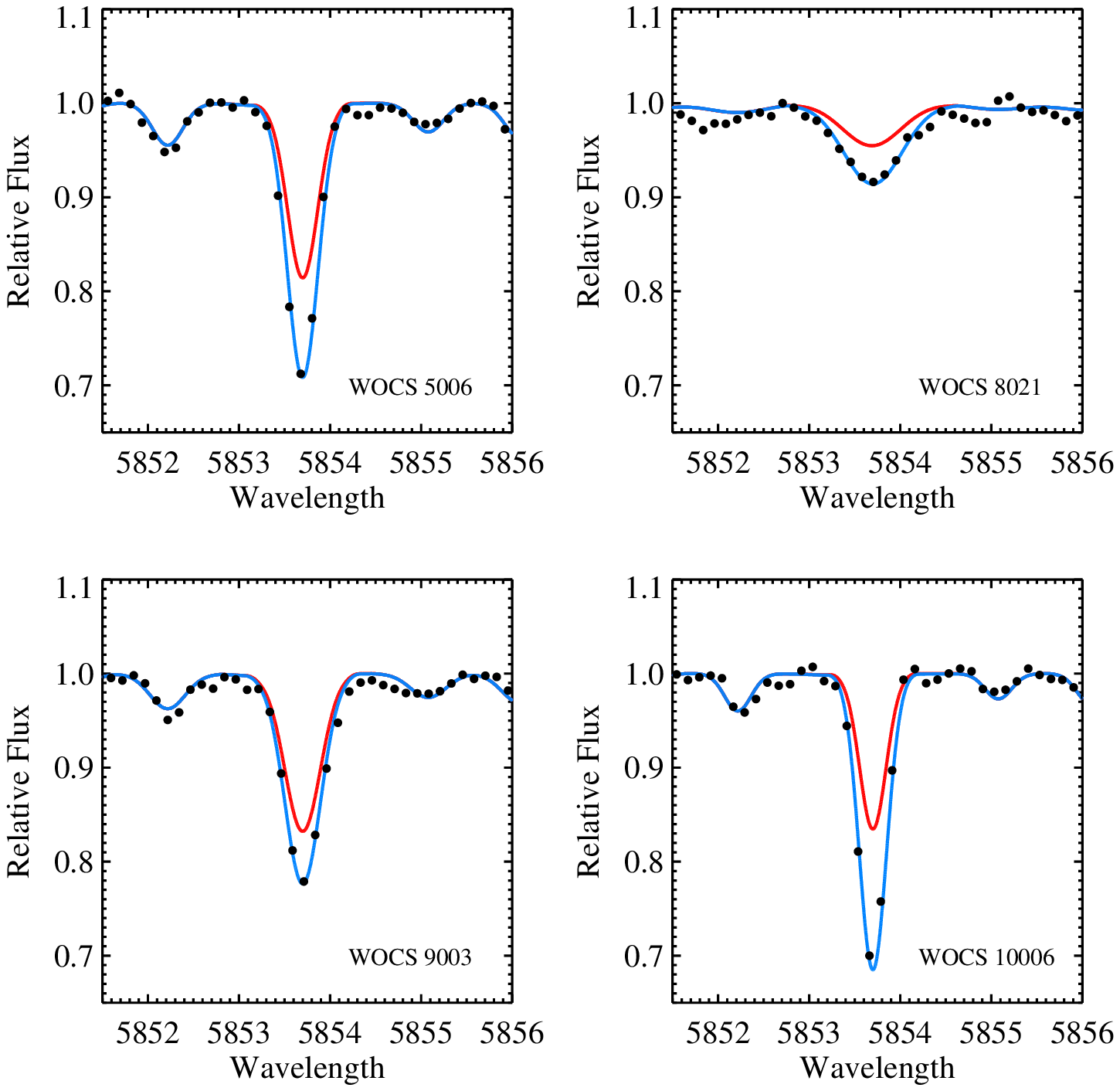}
\caption{The spectra for the blue stragglers with significant barium excess are shown (black dots). Overplotted are the computed profiles with [Ba/Fe] = 0.0 (red line) and the fitted barium abundance (blue line). }
\label{fig:bs}
\end{center}
\end{figure*}

As Figure~\ref{fig:abund} clearly shows, even allowing for considerable variations in temperature, gravity, and microturbulence, five BSs have barium abundances over the 3$\sigma$ variation of the MS value for NGC 6819: WOCS 5006, WOCS 8021, WOCS 9003, WOCS 10006, and the primary in WOCS 4004. These barium excesses are also evident in Figure~\ref{fig:bs} where we plot the observed spectra for four of these systems (see Figure~\ref{fig:4004} for the SB2, WOCS 4004). Along with our observed spectrum we plot two synthetic spectra with the atmospheric parameters of Table~\ref{tab:specs}; one has solar barium abundance and the other is our best fit spectrum that has significant barium enhancement.

\begin{deluxetable*}{lccrccrc}
\tablewidth{\textwidth}
\centering
\tabletypesize{\footnotesize}
\tablecaption{Abundance Information\label{tab:abund.info}}
\tablehead{\colhead{WOCS ID} & \colhead{} & \colhead{$N$} & \colhead{[Fe/H]} & \colhead{Range} & & \colhead{[Ba/Fe]}& \colhead{Range}}
\startdata
 1010 &      BS &  3 &     0.19 & \scriptsize{ $+$0.03 to $+$0.34 } &  &  $-$0.26 & \scriptsize{ $-$0.30 to $-$0.21} \\
 4004\tablenotemark{a} & BS & \nodata &    0.06 & \scriptsize{ $-$0.05 to $+$0.10 } &  &     1.36 & \scriptsize{ $+$1.10 to $+$1.50} \\
 5006 &      BS & 13 &     0.09 & \scriptsize{ $-$0.01 to $+$0.19 } &  &     0.94 & \scriptsize{ $+$0.84 to $+$1.16} \\
 8021 &      BS &  5 &     0.20 & \scriptsize{ $+$0.08 to $+$0.33 } &  &     0.57 & \scriptsize{ $+$0.42 to $+$0.67} \\
 9003 &      BS &  7 &  $-$0.06 & \scriptsize{ $-$0.16 to $+$0.04 } &  &     0.48 & \scriptsize{ $+$0.27 to $+$0.59} \\
10006 &      BS &  7 &     0.16 & \scriptsize{ $+$0.07 to $+$0.27 } &  &     0.99 & \scriptsize{ $+$0.62 to $+$1.14} \\
14015 &      BS &  3 &     0.03 & \scriptsize{ $-$0.06 to $+$0.13 } &  &  $-$0.14 & \scriptsize{ $-$0.22 to $-$0.07} \\
16009 &      BS &  6 &  $-$0.00 & \scriptsize{ $-$0.10 to $+$0.10 } &  &  $-$0.11 & \scriptsize{ $-$0.18 to $-$0.03} \\
16021 &      BS &  3 &  $-$0.01 & \scriptsize{ $-$0.11 to $+$0.10 } &  &     0.24 & \scriptsize{ $+$0.10 to $+$0.35} \\
21022 & \nodata &  4 &     0.06 & \scriptsize{ $-$0.05 to $+$0.19 } &  &  $-$0.12 & \scriptsize{ $-$0.22 to $-$0.02} \\
22018 & \nodata &  5 &     0.02 & \scriptsize{ $-$0.08 to $+$0.14 } &  &  $-$0.18 & \scriptsize{ $-$0.29 to $-$0.09} \\
22022 & \nodata &  8 &     0.06 & \scriptsize{ $-$0.04 to $+$0.18 } &  &     0.06 & \scriptsize{ $-$0.06 to $+$0.17} \\
23017 & \nodata &  7 &     0.07 & \scriptsize{ $-$0.03 to $+$0.20 } &  &  $-$0.11 & \scriptsize{ $-$0.21 to $-$0.03} \\
24027 & \nodata &  5 &     0.05 & \scriptsize{ $-$0.06 to $+$0.17 } &  &  $-$0.11 & \scriptsize{ $-$0.20 to $-$0.04} \\
25027 & \nodata &  5 &     0.02 & \scriptsize{ $-$0.08 to $+$0.13 } &  &     0.15 & \scriptsize{ $+$0.01 to $+$0.24} \\
26017 & \nodata &  3 &     0.07 & \scriptsize{ $-$0.04 to $+$0.19 } &  &     0.02 & \scriptsize{ $-$0.09 to $+$0.12} \\
26019 & \nodata &  5 &     0.01 & \scriptsize{ $-$0.09 to $+$0.12 } &  &  $-$0.08 & \scriptsize{ $-$0.17 to $-$0.02} \\
27019 & \nodata &  8 &     0.04 & \scriptsize{ $-$0.06 to $+$0.16 } &  &  $-$0.05 & \scriptsize{ $-$0.14 to $+$0.05} \\
27021 & \nodata &  7 &     0.00 & \scriptsize{ $-$0.10 to $+$0.12 } &  &     0.01 & \scriptsize{ $-$0.11 to $+$0.09} \\
28017 & \nodata &  8 &     0.07 & \scriptsize{ $-$0.03 to $+$0.18 } &  &     0.03 & \scriptsize{ $-$0.10 to $+$0.13} \\
30027 & \nodata &  6 &     0.03 & \scriptsize{ $-$0.08 to $+$0.15 } &  &     0.06 & \scriptsize{ $-$0.06 to $+$0.17} \\
31028 & \nodata & 11 &  $-$0.02 & \scriptsize{ $-$0.12 to $+$0.10 } &  &     0.07 & \scriptsize{ $-$0.04 to $+$0.18} \\
32023 &      BS &  5 &     0.05 & \scriptsize{ $-$0.04 to $+$0.15 } &  &  $-$0.00 & \scriptsize{ $-$0.07 to $+$0.07} \\
32026 & \nodata &  6 &     0.05 & \scriptsize{ $-$0.05 to $+$0.17 } &  &  $-$0.03 & \scriptsize{ $-$0.13 to $+$0.06} \\
33038 & \nodata & 11 &     0.09 & \scriptsize{ $-$0.02 to $+$0.20 } &  &  $-$0.16 & \scriptsize{ $-$0.26 to $-$0.08} \\
41039 & \nodata &  5 &     0.11 & \scriptsize{ $+$0.01 to $+$0.23 } &  &     0.09 & \scriptsize{ $-$0.04 to $+$0.20} \\
47030 &      BS &  4 &     0.00 & \scriptsize{ $-$0.09 to $+$0.11 } &  &  $-$0.03 & \scriptsize{ $-$0.10 to $+$0.04} \\
51059 &      BS &  6 &  $-$0.04 & \scriptsize{ $-$0.14 to $+$0.07 } &  &  $-$0.01 & \scriptsize{ $-$0.08 to $+$0.08} \\
62034 & \nodata &  3 &     0.01 & \scriptsize{ $-$0.10 to $+$0.14 } &  &     0.07 & \scriptsize{ $-$0.06 to $+$0.18} \\
63044 & \nodata &  8 &  $-$0.06 & \scriptsize{ $-$0.16 to $+$0.07 } &  &  $-$0.14 & \scriptsize{ $-$0.22 to $-$0.05} 
\enddata
\tablenotetext{a}{WOCS ID 4004 is a double-lined system. The abundance values for the primary are listed.}
\end{deluxetable*}

%Figure 3; abund.ps

%Figure 4; spec.bs.ps

\section{Abundance Discussion}
These barium excesses indicate pollution of these five BS surfaces by an AGB companion and a mass transfer origin for these systems, but there are some intriguing inconsistencies with this narrative that we explore below.

\subsection{Binarity of Barium-enhanced Systems}
\label{sec:binarity}

The standard threshold used in WOCS RV surveys to determine if a star is velocity-variable is $e$/$i$ $>$ 4.0. However, $e$/$i$ has a value less than 2.0 for all but one barium-enriched BS. With over 24 years of RV observations these four low $e$/$i$ targets are firmly categorized as non-velocity-variable. The one star that does have RV variation, WOCS 4004, has a companion that is definitively not a white dwarf and shows no means of providing the significant amount of barium measured on the surface of the BS. The current narrative of mass transfer from an evolved companion forming these BS systems, polluting the BS's surface with barium and leaving behind a degenerate companion, does not evidently explain these systems.

We suggest that the double-lined system, WOCS 4004, may have undergone a dynamical exchange after BS formation resulting in the current system with a luminous secondary. The normal abundance values of the secondary is consistent with this scenario. It is extremely unlikely that all five of these enhanced systems underwent dynamical encounters after BS formation resulting in four non-velocity variable stars and one double-lined binary. Especially since dynamical interactions for binaries with periods under 10$^{6}$ days typically result in shorter-period, tighter orbits (\citealt{Heggie1975}). With no other mechanism available to produce the barium abundances observed besides an AGB companion and the low probability of the dynamical destruction of all these binaries, we look to other means of explaining these results, none of which offers a completely satisfactory solution. 

One possibility is that the four enhanced systems with no RV variation have companions that we are not able to detect. We modeled this scenario with the same method used to determine the overall binary detection completeness of the WOCS RV survey outlined in \cite{Geller2012} and detailed for NGC 6819 in \cite{Milliman2014}. Specifically, we generated artificial binaries with period and eccentricity distributions of the Galactic field binary population found by \cite{R2010}. We assumed a secondary white dwarf mass of 0.6 \Msolar~and a primary mass that ranged from 1.6 \Msolar~to 2.4 \Msolar. Orbital inclination, longitude of periastron, and phase were chosen randomly. We used the actual observation dates and measurement precisions in the model. RV measurements were produced for a large number of synthetic binaries and their detectability assessed.
Our detection completeness depends most strongly on the period of the binary system and has little dependence on the mass ratio. For each mass ratio our detection capability falls below 50\% for periods over 15,500 days and below 25\% for periods over 40,000 days. 

Such long periods severely restrict the amount of mass that can be transferred to a companion. Adjusting parameters within the Binary-stellar Evolution code (BSE) from \cite{Hurley2002}, the maximum mass we are able to transfer that resulted in a binary orbit over 15,500 days was under 0.12 \Msolar. This value required highly eccentric systems ($e$ $\geqslant$ 0.5).

A similar low mass enhancement was also suggested by \cite{Gosnell2014} for a long-period (3030 $\pm$ 70 days) BS with a detected white dwarf companion in NGC 188. Using BSE, they found the accretor received 0.101 \Msolar~through wind accretion. Surprisingly, this BS is one of the most luminous BS binaries in NGC 188. As mentioned in \cite{Gosnell2014} this raises many questions about the mass-luminosity relation in BS stars, whether wind accretion is the correct scenario, and if the AGB wind prescription used in BSE is correct. The formation of these four potentially very long period barium enhanced stars adds to these questions. 

Another explanation for these non-velocity variable stars arises from work done with barium stars, red giants with enhanced carbon and $s$-process elements. Binary population synthesis studies including \cite{Pols2003} and \cite{Izzard2010} seek to explain the origin of barium stars and match the observed population fraction, measured overabundances, and the distribution of periods and eccentricities of these mostly binary systems. Without modified physics these studies do not reproduce the observed barium star period-eccentricity distribution. Specifically they produce no eccentric systems with $P$ $\lesssim$ 3000 days and generate too many long period ($P$ $\gtrsim$ 10$^{4}$) barium stars. In order to address this, \cite{Izzard2010} apply a white-dwarf kick in their model as a way to generate the eccentricities observed in binary barium stars (although the mechanism for producing the white dwarf kick is unknown).
 
In the model this white dwarf kick also pushes weakly bound, long-period systems into wider orbits and even disrupts some binary systems. Either of these two scenarios could explain the non-velocity-variable enhanced systems seen in NGC 6819. Unfortunately observations of actual barium stars rule out a large population of long period systems and the overwhelmingly large fraction of barium stars detected in binaries means a very small disruption rate for these systems (\citealt{Jorissen1998}), so a strong white dwarf kick seems unlikely. 

It is unclear why, if these enriched open cluster BSs and field barium stars all formed through the same process, their orbital properties would be so different. Hopefully more complete binary evolution and $N$-body modeling for NGC 6819 will provide some insight.

\subsection{Amount of Barium Enhancement}
The five barium-enhanced BSs have [Ba/Fe] ranging from +0.48 (WOCS 9003) to +1.36 (WOCS 4004). We have suggested that the origin of the excess barium on these BSs is the $s$-process element production that happened in the interior of AGB stars and subsequently was mass transferred to the BSs. Here we address whether these levels of barium enhancement are feasible. 

First we assume that the AGB companion evolved as it would in isolation. This is appropriate because, as discussed in section~\ref{sec:binarity}, if these non-velocity-variable BS with barium excess are in binaries they have long-periods and the binary would not have significantly disturbed the stellar evolution, thermal pulses, or barium production of any AGB companion. 
 
We also assume that an accreting star would have a mass near the turnoff mass of NGC 6819, $\sim$1.5 \Msolar, and would not have efficient surface convection zones that would significantly dilute the barium abundance of any accreted material. This is supported by MESA (\citealt{Paxton2011}) models of solar metallicity stars, where a 1.4 \Msolar~and 1.2 \Msolar~star have only $\sim$0.0006 \Msolar~and 0.0015 \Msolar, respectively, in a surface convection zone at an age of 2.5 Gyr.  

The accreted material would have a higher mean molecular weight than the surface layer of the accretor and thermohaline mixing should eliminate the molecular weight difference on thermal time-scales (\citealt{Kippenhahn1980}). However, abundance measurements of carbon-enhanced metal-poor stars have called into question the efficiency of thermohaline mixing and various reduction scenarios have been investigated (\citealt{Aoki2008}; \citealt{Stancliffe2008}; \citealt{Thompson2008}; \citealt{Bisterzo2011}). Based on this we assume no significant amount of dilution of the barium surface abundance has taken place since the time the enriched material was accreted.

To compare our observational results with theory we use \cite{Cristallo2011} models of the nucleosynthesis processes of AGB stars of different masses. These results are available from the FRUITY Database\footnote{http://fruity.oa-teramo.inaf.it/} that provides the isotopic composition from H to Bi as it changes after each third dredge-up (TDU) episode and the stellar yields produced for each model. \textit{Kepler} asteroseismology puts the average mass of red clump stars in NGC 6819 at 1.64 \Msolar~(\citealt{Miglio2011}). The FRUITY database has models for 1.5 \Msolar~and 2.0 \Msolar. Assuming solar metallicity ($z$ = 0.014) and no initial rotational velocity, the 1.5 \Msolar~model experiences 5 TDU episodes with a final [Ba/Fe] = $+$0.65 and the 2.0 \Msolar~model undergoes 9 TDU events with a final [Ba/Fe] = $+$1.03.

All of this indicates that the amount of barium produced by a $\sim$1.5 \Msolar~to $\sim$2 \Msolar~initial-mass AGB companion can broadly account for the amount of barium excess seen in these NGC 6819 BSs.

\subsection{Age and Position in the CMD}
As outlined in the previous section, the levels of barium excess for the three stars with the highest enrichment (WOCS 4004, WOCS 5006, and WOCS 10006) require a donor with $\sim$2 \Msolar~initial mass. Based on the 1.5 Gyr evolutionary lifetime of a 2.0 \Msolar~star (from the MS through the TP-AGB; \citealt{PARSEC2012}) and the 2.5 Gyr cluster age of NGC 6819 that places the formation of these BSs at roughly 1 Gyr ago. 

The barium enhancement levels of the other excess BSs (WOCS 8021 and WOCS 9003) are more in line with the expected [Ba/Fe] levels of the current AGB stars at 1.64 \Msolar~(\citealt{Miglio2011}). So we expect these BSs to be young systems and to have ages much less than 1 Gyr, but we are unable to quantify when they formed.  

\begin{figure}[htbp]
\begin{center} 
\includegraphics[width=0.95\linewidth]{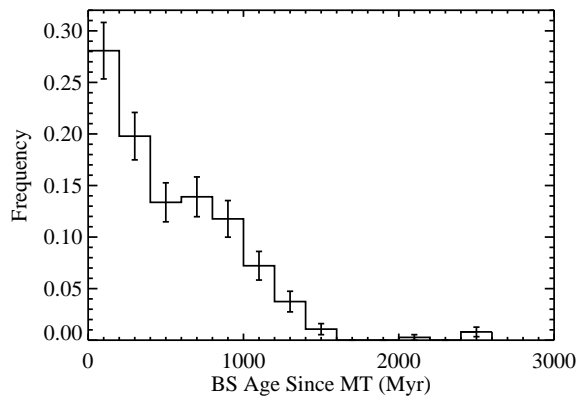}
\caption{Histogram of the ages for mass transfer-formed BSs from the NGC 188 $N$-body model of \cite{Geller2013} from when the cluster was 2$-$3 Gyr.}
\label{fig:bs.ages}
\end{center}
\end{figure}

We can compare the age estimates of $\sim$1 Gyr to the sophisticated $N$-body models of \cite{Geller2013} for NGC 188. These models track BS formation throughout the cluster's lifetime up to the present NGC 188 age of 7 Gyr. Taking snapshots of the 20 realizations of the NGC 188 model at a cluster age between 2 and 3 Gyr, we construct the age distribution of mass transfer-formed BSs (i.e., the time since the end of mass transfer) that we expect in a cluster like NGC 6819. We plot this age distribution in Figure~\ref{fig:bs.ages} and show that at a cluster age of 2$-$3 Gyr the mass transfer-formed BSs are typically young systems, but there is a long tail that extends to roughly 1.6 Gyr. Our Monte Carlo-sampling of this distribution shows that finding three or more BSs with ages older than 1 Gyr is not restrictively uncommon, happening $\sim$20\% of the time when we assume a total of 12 mass transfer-formed BSs. 

These three BS age estimates of $\sim$1 Gyr are much older than the seven other known open cluster BS ages that range from 70 Myr to 400 Myr. These other BS ages were all determined for BSs in the old open cluster NGC 188 by \cite{GosnellPhD}. The \cite{GosnellPhD} study was designed to find the most recently formed mass transfer BSs through detection of the UV excess produced by hot, young white dwarf companions. On the other hand our study determines age estimates for only the most barium rich and oldest BSs. So we consider these different age determinations to be a consequence of the observational biases of the two surveys. 

%Figure 5; bs.ages.eps

\begin{figure}[htbp]
\begin{center} 
\includegraphics[width=\linewidth]{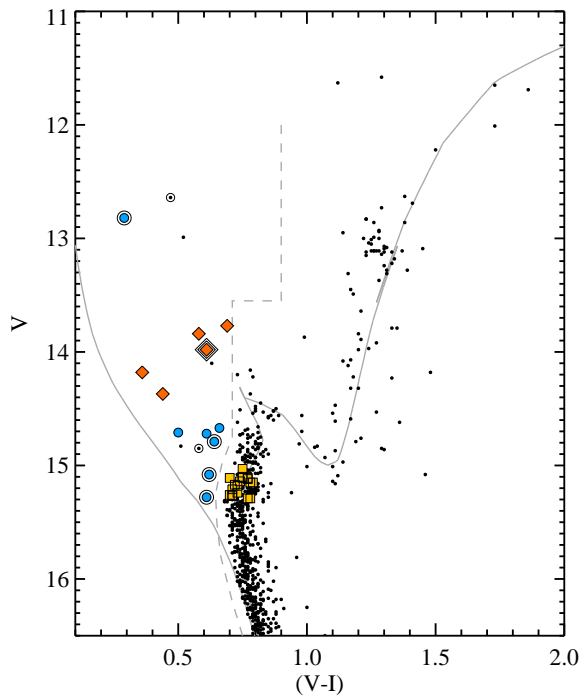}
\caption{Same as Fig.~\ref{fig:cmd} except BSs with significant barium excess are marked by red diamonds.}
\label{fig:cmd.excess}
\end{center}
\end{figure}

We highlight the CMD position of the five NGC 6819 BSs with significant barium excess in Figure~\ref{fig:cmd.excess}. All five BSs occupy a narrow region and fall between the evolutionary tracks of a $\sim$1.75 \Msolar~and a $\sim$1.95 \Msolar~star (\citealt{PARSEC2012}). We do not make mass estimates for these stars, because it is not clear if these standard tracks are appropriate for BS mass estimates (\citealt{MathieuGeller2015}), but this positioning would suggest that these BSs have similar masses. The barium excess systems have varying distances from the zero-age main-sequence (ZAMS) which might be attributable to age differences. However, their positions have no apparent dependence on barium abundance values. The two BS closest to the ZAMS are WOCS 8021, [Ba/Fe] = $+$0.57 and WOCS 10006, [Ba/Fe] = $+$0.99. The three furthest from the ZAMS are the primary of WOCS 4004\footnote{This order is based on the expected position of the primary from a simplistic deconvolution of the SB2 photometry.}, [Ba/Fe] = $+$1.36; WOCS 5006, [Ba/Fe] = $+$0.94; and finally WOCS 9003, [Ba/Fe] = $+$0.48. This distribution shows that the most barium-enhanced, and presumably oldest BSs at $\sim$1 Gyr, are not the farthest from the ZAMS.

If in fact all of these BSs have comparable masses, these results suggest that barium abundance or CMD position are not accurate measures for BS age. Additional evidence that BS age and CMD position is not a straight-forward relationship comes from NGC 188. \cite{GosnellPhD} found that the distance from the ZAMS is not necessarily equivalent with BS age. This conclusion was based on the positions off the ZAMS of the definitively very young BSs in NGC 188. \cite{GosnellPhD} suggests that BS ages based on single-star isochrones are unreliable. Our results also seem to suggest this although this conclusion is clouded by our assumptions.

As for why the barium-enhanced systems in NGC 6819 occupy such a narrow region in the optical CMD, we have no satisfying explanation. 

%Figure 6; cmd.excess.eps 

\section{Formation Mechanisms for the Other Blue Stragglers}
The origin of the other 7 NGC 6819 BSs in this abundance survey is still an open question. The lack of significant barium enhancement itself does not rule out a mass-transfer formation history for a BS. If the BS progenitor binary had a close enough orbit, the mass transfer event could have occurred before the giant companion made it to the AGB phase, or the binary could have stripped enough of the AGB envelope to prevent thermal pulses and any substantial barium production. Future work will focus on other abundance tracers that are indicative of earlier mass transfer activity. We will particularly focus on oxygen and carbon anomalies that would indicate that mass transfer originated along the red giant branch as seen in some globular cluster BSs (e.g., \citealt{Ferraro2006}; \citealt{Lovisi2013M30}). 

It is harder to explain the lack of barium enrichment for long-period BS binaries. WOCS 1010 has an orbital period of 1,144 days, an eccentricity of 0.55, and has the lowest [Ba/Fe] of any of the BS surveyed. WOCS 1010 is also the hottest BS in NGC 6819 with a temperature of 8524 K, meaning its surface abundance is most likely strongly affected by radiative levitation (\citealt{Lovisi2012}, \citealt{Lovisi2013}) and we are unable to draw any solid conclusions from the abundance results. 

Another long period BS binary is WOCS 47030. It has a \teff~= 6820 K and does not have a completed orbital solution, but the RV data suggest an eccentric orbit ($e$ $\sim$ 0.7) with a period around 3700 days. At that period the most likely mass transfer formation scenario would involve wind accretion from an AGB companion, but no enrichment in [Ba/Fe] is measured. This leaves open the possibility that WOCS 47030 formed from the merger of a hierarchical triple system as described in \cite{Perets2009}.

We note that although \cite{GellerNature2011} and \cite{GosnellPhD} establish mass transfer to be the dominant BS formation process in NGC 188 and we find at least $\sim$30\% (5/17) of the BS in NGC 6819 to be formed from specifically AGB mass transfer events, we still expect collisions and mergers to have formed some BSs in NGC 6819. 

\section{Summary and Discussion} 
As a way to distinguish between various BS formation mechanisms we measured the abundance of the $s$-process element barium in 12 BSs in the intermediate-age open cluster NGC 6819. Our observations of these BSs and 18 MS stars were made in 2014 July using the Hydra Multi-object Spectrograph on the WIYN 3.5 m telescope. 

For our control sample of MS stars, we find an average [Fe/H] = +0.05 $\pm$ 0.04 and [Ba/Fe] = $-$0.01 $\pm$ 0.10. This iron abundance is consistent with previous spectroscopic abundance studies of NGC 6819. The average [Fe/H] value for the BSs is also consistent, [Fe/H] = +0.06 $\pm$ 0.09. 

Out of the 12 BSs in our sample we identify five that have significant barium enrichment. From the amount of barium enhancement on these five BSs we find that:
\begin{enumerate}
\item A $\sim$1.5 \Msolar~to $\sim$2.0 \Msolar~initial-mass AGB companion is able to produce that amount of [Ba/Fe] measured.
\item This AGB mass range matches well with the current NGC 6819 red clump mass determined from \textit{Kepler} observations.
\item The three most enriched BSs formed roughly 1 Gyr ago.
\end{enumerate}

We suggest that these barium enhanced systems formed via mass transfer from an AGB binary companion that polluted the BS with the nucleosynthesis products of the AGB star. This formation scenario is complicated by the fact that:
\begin{enumerate}
\item Four of the enhanced systems show no RV evidence for a companion.
\item If these non-velocity-variable systems are indeed in long-period binaries, we place the expected orbital periods at over 15,500 days.
\item Such wide orbits severely limits the amount of mass that can be transferred to a companion and we find the maximum amount transferred to be under 0.12 \Msolar. 
\item The one enriched system in a binary has a luminous companion with no means of providing the excess barium observed. 
\item This proposed formation scenario is similar to how barium stars form, but we find the barium-enriched BSs in NGC 6819 have very different binary properties than these enhanced field giants.
\end{enumerate}

Future work should focus on incorporating more detailed binary evolution modeling for these barium enriched BSs, perhaps with a different AGB wind prescription, and $N$-body models specifically for NGC 6819 to help explain the origin of these enriched systems and their lack of RV-detected companions. Additional abundance tracers, such as other neutron capture elements and carbon, may also shed light on the formation histories of these enhanced BSs, as well as the other BSs in NGC 6819.

Moving beyond NGC 6819, we predict that there is an upper limit on cluster age for finding barium enhanced BSs formed via mass transfer from an AGB binary companion. The FRUITY database has no TDU episodes or $s$-process element production occurring for a solar-metallicity 1.3 \Msolar~AGB star. Presumably this is because mass loss has reduced the envelope mass below the $\sim$0.4$-$0.5 \Msolar~threshold required for TDU to occur (\citealt{Cristallo2011}). We estimate this barium-enhancement cluster age limit by combining the standard stellar lifetime of a 1.5 \Msolar~star (the lowest mass model on the FRUITY database that produces barium) and the amount of time a star ``survives'' as a BS. We use a PARSEC evolutionary track to get a standard stellar lifetime of 3 Gyr. The tail of the $N$-body mass transfer-formed BS age distribution of \cite{Geller2013} provides us with $\sim$2 Gyr as the typical upper end of BS survival time.  So we place this cluster age limit at roughly 5 Gyr and do not expect barium-enhanced BSs in open clusters\footnote{Lower metallicity environments will have a different age limit.} older than 5 Gyr. Indeed, the surface abundance study of the BSs in the 7 Gyr open cluster NGC 188 found no obvious barium enhancements (\citealt{Milliman2013}). 

We recommend that future BS abundance surveys look for $s$-process element enhancements in intermediate-aged open clusters younger than NGC 6819, but we do caution that clusters that are too young will potentially have more high temperature BSs affected by radiative levitation.\\

The authors express their thanks to Dr. Aaron Geller for his valuable input and helpful discussion regarding $N$-body models and blue straggler ages. This work was funded by the National Science Foundation grant AST-0908082 to the University of Wisconsin-Madison and several awards from the Graduate School of the University of Wisconsin-Madison. K. E. M. is supported in part by the National Space Grant College and Fellowship Program and the Wisconsin Space Grant Consortium. 

\bibliographystyle{mn2e}
\bibliography{abund}

\begin{thebibliography}{56}
\expandafter\ifx\csname natexlab\endcsname\relax\def\natexlab#1{#1}\fi

\bibitem[{{Anthony-Twarog} {et~al}\mbox{.}(2013){Anthony-Twarog}, {Deliyannis},
  {Rich}, \& {Twarog}}]{AnthonyTwarog2013}
{Anthony-Twarog} B.~J., {Deliyannis} C.~P., {Rich} E., {Twarog} B.~A., 2013,
  \apjl, 767, L19

\bibitem[{{Anthony-Twarog} {et~al}\mbox{.}(2010){Anthony-Twarog}, {Deliyannis},
  {Twarog}, {et~al.}}]{AnthonyTwarog2010}
{Anthony-Twarog} B.~J., {Deliyannis} C.~P., {Twarog} B.~A., {et~al.}, 2010,
  \aj, 139, 2034

\bibitem[{{Aoki} {et~al}\mbox{.}(2008){Aoki}, {Beers}, {Sivarani},
  {et~al.}}]{Aoki2008}
{Aoki} W., {Beers} T.~C., {Sivarani} T., {et~al.}, 2008, \apj, 678, 1351

\bibitem[{{Barden} {et~al}\mbox{.}(1994){Barden}, {Armandroff}, {Muller},
  {et~al.}}]{Barden1994}
{Barden} S.~C., {Armandroff} T., {Muller} G., {et~al.}, 1994, in Proc. SPIE,
  Vol. 2198, Instrumentation in Astronomy VIII, {Crawford} D.~L., {Craine}
  E.~R., eds., pp. 87--97

\bibitem[{{Bisterzo} {et~al}\mbox{.}(2011){Bisterzo}, {Gallino}, {Straniero},
  {et~al.}}]{Bisterzo2011}
{Bisterzo} S., {Gallino} R., {Straniero} O., {et~al.}, 2011, \mnras, 418, 284

\bibitem[{{Blackwell} \& {Lynas-Gray}(1998)}]{BLG1998}
{Blackwell} D.~E., {Lynas-Gray} A.~E., 1998, \aaps, 129, 505

\bibitem[{{Bragaglia} {et~al}\mbox{.}(2001){Bragaglia}, {Carretta}, {Gratton},
  {et~al.}}]{Brag2001}
{Bragaglia} A., {Carretta} E., {Gratton} R.~G., {et~al.}, 2001, \aj, 121, 327

\bibitem[{{Bressan} {et~al}\mbox{.}(2012){Bressan}, {Marigo}, {Girardi},
  {et~al.}}]{PARSEC2012}
{Bressan} A., {Marigo} P., {Girardi} L., {et~al.}, 2012, \mnras, 427, 127

\bibitem[{{Carlberg} {et~al}\mbox{.}(2015){Carlberg}, {Smith}, {Cunha},
  {Majewski}, {M{\'e}sz{\'a}ros}, {Shetrone}, {Allende Prieto}, {Bizyaev},
  {Stassun}, {Fleming}, {Zasowski}, {Hearty}, {Nidever}, {Schneider},
  {Holtzman}, \& {Frinchaboy}}]{Carlberg2015}
{Carlberg} J.~K. {et~al.}, 2015, \apj, 802, 7

\bibitem[{{Castelli} \& {Kurucz}(2004)}]{ATLAS9}
{Castelli} F., {Kurucz} R.~L., 2004, ArXiv Astrophysics e-prints

\bibitem[{{Corsaro} {et~al}\mbox{.}(2012){Corsaro}, {Stello}, {Huber},
  {et~al.}}]{Corsaro2012}
{Corsaro} E., {Stello} D., {Huber} D., {et~al.}, 2012, \apj, 757, 190

\bibitem[{{Cristallo} {et~al}\mbox{.}(2011){Cristallo}, {Piersanti},
  {Straniero}, {et~al.}}]{Cristallo2011}
{Cristallo} S., {Piersanti} L., {Straniero} O., {et~al.}, 2011, \apjs, 197, 17

\bibitem[{{Edvardsson} {et~al}\mbox{.}(1993){Edvardsson}, {Andersen},
  {Gustafsson}, {et~al.}}]{Edvardsson1993}
{Edvardsson} B., {Andersen} J., {Gustafsson} B., {et~al.}, 1993, \aap, 275, 101

\bibitem[{{Ferraro} {et~al}\mbox{.}(2006){Ferraro}, {Sabbi}, {Gratton},
  {et~al.}}]{Ferraro2006}
{Ferraro} F.~R., {Sabbi} E., {Gratton} R., {et~al.}, 2006, \apjl, 647, L53

\bibitem[{{Geller}, {Hurley} \& {Mathieu}(2013){Geller}, {Hurley}, \&
  {Mathieu}}]{Geller2013}
{Geller} A.~M., {Hurley} J.~R., {Mathieu} R.~D., 2013, \aj, 145, 8

\bibitem[{{Geller} \& {Mathieu}(2011)}]{GellerNature2011}
{Geller} A.~M., {Mathieu} R.~D., 2011, \nat, 478, 356

\bibitem[{{Geller} \& {Mathieu}(2012)}]{Geller2012}
{Geller} A.~M., {Mathieu} R.~D., 2012, \aj, 144, 54

\bibitem[{{Geller} {et~al}\mbox{.}(2008){Geller}, {Mathieu}, {Harris}, \&
  {McClure}}]{Geller2008}
{Geller} A.~M., {Mathieu} R.~D., {Harris} H.~C., {McClure} R.~D., 2008, \aj,
  135, 2264

\bibitem[{{Gosnell}(2014)}]{GosnellPhD}
{Gosnell} N.~M., 2014, PhD thesis, The University of Wisconsin - Madison

\bibitem[{{Gosnell} {et~al}\mbox{.}(2014){Gosnell}, {Mathieu}, {Geller},
  {et~al.}}]{Gosnell2014}
{Gosnell} N.~M., {Mathieu} R.~D., {Geller} A.~M., {et~al.}, 2014, \apjl, 783,
  L8

\bibitem[{{Gosnell} {et~al}\mbox{.}(2012){Gosnell}, {Pooley}, {Geller},
  {et~al.}}]{Gosnell2012}
{Gosnell} N.~M., {Pooley} D., {Geller} A.~M., {et~al.}, 2012, \apj, 745, 57

\bibitem[{{Heggie}(1975)}]{Heggie1975}
{Heggie} D.~C., 1975, \mnras, 173, 729

\bibitem[{{Hole} {et~al}\mbox{.}(2009){Hole}, {Geller}, {Mathieu},
  {et~al.}}]{Hole2009}
{Hole} K.~T., {Geller} A.~M., {Mathieu} R.~D., {et~al.}, 2009, \aj, 138, 159

\bibitem[{{Hurley}, {Tout} \& {Pols}(2002){Hurley}, {Tout}, \&
  {Pols}}]{Hurley2002}
{Hurley} J.~R., {Tout} C.~A., {Pols} O.~R., 2002, \mnras, 329, 897

\bibitem[{{Izzard}, {Dermine} \& {Church}(2010){Izzard}, {Dermine}, \&
  {Church}}]{Izzard2010}
{Izzard} R.~G., {Dermine} T., {Church} R.~P., 2010, \aap, 523, A10

\bibitem[{{Jorissen} {et~al}\mbox{.}(1998){Jorissen}, {Van Eck}, {Mayor}, \&
  {Udry}}]{Jorissen1998}
{Jorissen} A., {Van Eck} S., {Mayor} M., {Udry} S., 1998, \aap, 332, 877

\bibitem[{{Kippenhahn}, {Ruschenplatt} \& {Thomas}(1980){Kippenhahn},
  {Ruschenplatt}, \& {Thomas}}]{Kippenhahn1980}
{Kippenhahn} R., {Ruschenplatt} G., {Thomas} H.-C., 1980, \aap, 91, 175

\bibitem[{{Korotin} {et~al}\mbox{.}(2011){Korotin}, {Mishenina}, {Gorbaneva},
  \& {Soubiran}}]{Korotin2011}
{Korotin} S., {Mishenina} T., {Gorbaneva} T., {Soubiran} C., 2011, \mnras, 415,
  2093

\bibitem[{{Kupka} {et~al}\mbox{.}(2000){Kupka}, {Ryabchikova}, {Piskunov},
  {et~al.}}]{VALD2000}
{Kupka} F.~G., {Ryabchikova} T.~A., {Piskunov} N.~E., {et~al.}, 2000, BaltA, 9,
  590

\bibitem[{{Lee-Brown} {et~al}\mbox{.}(2015){Lee-Brown}, {Anthony-Twarog},
  {Deliyannis}, {et~al.}}]{LeeBrown2015}
{Lee-Brown} D.~B., {Anthony-Twarog} B.~J., {Deliyannis} C.~P., {et~al.}, 2015,
  \aj, 149, 121

\bibitem[{{Lovisi} {et~al}\mbox{.}(2013{\natexlab{a}}){Lovisi}, {Mucciarelli},
  {Dalessandro}, {et~al.}}]{Lovisi2013}
{Lovisi} L., {Mucciarelli} A., {Dalessandro} E., {et~al.}, 2013{\natexlab{a}},
  \apj, 778, 64

\bibitem[{{Lovisi} {et~al}\mbox{.}(2012){Lovisi}, {Mucciarelli}, {Lanzoni},
  {et~al.}}]{Lovisi2012}
{Lovisi} L., {Mucciarelli} A., {Lanzoni} B., {et~al.}, 2012, \apj, 754, 91

\bibitem[{{Lovisi} {et~al}\mbox{.}(2013{\natexlab{b}}){Lovisi}, {Mucciarelli},
  {Lanzoni}, {et~al.}}]{Lovisi2013M30}
{Lovisi} L., {Mucciarelli} A., {Lanzoni} B., {et~al.}, 2013{\natexlab{b}},
  \apj, 772, 148

\bibitem[{{Mathieu}(2000)}]{Mathieu2000}
{Mathieu} R.~D., 2000, in ASP Conf. Ser., Vol. 198, Stellar Clusters and
  Associations: Convection, Rotation, and Dynamos, {Pallavicini} R., {Micela}
  G., {Sciortino} S., eds., p. 517

\bibitem[{{Mathieu} \& {Geller}(2009)}]{MathieuNature2009}
{Mathieu} R.~D., {Geller} A.~M., 2009, \nat, 462, 1032

\bibitem[{{Mathieu} \& {Geller}(2015)}]{MathieuGeller2015}
{Mathieu} R.~D., {Geller} A.~M., 2015, {The Blue Stragglers of the Old Open
  Cluster NGC 188}, {Boffin} H.~M.~J., {Carraro} G., {Beccari} G., eds., Vol.
  413, Springer-Verlag Berlin Heidelberg, p.~29

\bibitem[{{McGahee} {et~al}\mbox{.}(2013){McGahee}, {King}, {Deliyannis}, \&
  {Maderak}}]{McGahee2013}
{McGahee} C., {King} J.~R., {Deliyannis} C.~P., {Maderak} R.~M., 2013, in
  American Astronomical Society Meeting Abstracts, Vol. 222, American
  Astronomical Society Meeting Abstracts, p. 116.09

\bibitem[{{Miglio} {et~al}\mbox{.}(2012){Miglio}, {Brogaard}, {Stello},
  {et~al.}}]{Miglio2011}
{Miglio} A., {Brogaard} K., {Stello} D., {et~al.}, 2012, \mnras, 419, 2077

\bibitem[{{Milliman}, {Mathieu} \& {Schuler}(2013){Milliman}, {Mathieu}, \&
  {Schuler}}]{Milliman2013}
{Milliman} K., {Mathieu} R.~D., {Schuler} S.~C., 2013, in BASS, Vol. 222,
  American Astronomical Society Meeting Abstracts, p. 214.05

\bibitem[{{Milliman} {et~al}\mbox{.}(2014){Milliman}, {Mathieu}, {Geller},
  {et~al.}}]{Milliman2014}
{Milliman} K.~E., {Mathieu} R.~D., {Geller} A.~M., {et~al.}, 2014, \aj, 148, 38

\bibitem[{{Paxton} {et~al}\mbox{.}(2011){Paxton}, {Bildsten}, {Dotter},
  {et~al.}}]{Paxton2011}
{Paxton} B., {Bildsten} L., {Dotter} A., {et~al.}, 2011, \apjs, 192, 3

\bibitem[{{Perets} \& {Fabrycky}(2009)}]{Perets2009}
{Perets} H.~B., {Fabrycky} D.~C., 2009, \apj, 697, 1048

\bibitem[{{Platais} {et~al}\mbox{.}(2013){Platais}, {Gosnell}, {Meibom},
  {et~al.}}]{Platais2013}
{Platais} I., {Gosnell} N.~M., {Meibom} S., {et~al.}, 2013, \aj, 146, 43

\bibitem[{{Pols} {et~al}\mbox{.}(2003){Pols}, {Karakas}, {Lattanzio}, \&
  {Tout}}]{Pols2003}
{Pols} O.~R., {Karakas} A.~I., {Lattanzio} J.~C., {Tout} C.~A., 2003, in ASP
  Conf. Ser., Vol. 303, Symbiotic Stars Probing Stellar Evolution, {Corradi}
  R.~L.~M., {Mikolajewska} J., {Mahoney} T.~J., eds., p. 290

\bibitem[{{Raghavan} {et~al}\mbox{.}(2010){Raghavan}, {McAlister}, {Henry},
  {et~al.}}]{R2010}
{Raghavan} D., {McAlister} H.~A., {Henry} T.~J., {et~al.}, 2010, \apjs, 190, 1

\bibitem[{{Ram{\'{\i}}rez}, {Allende Prieto} \&
  {Lambert}(2013){Ram{\'{\i}}rez}, {Allende Prieto}, \&
  {Lambert}}]{Ramirez2013}
{Ram{\'{\i}}rez} I., {Allende Prieto} C., {Lambert} D.~L., 2013, \apj, 764, 78

\bibitem[{{Rosvick} \& {Vandenberg}(1998)}]{RV1998}
{Rosvick} J.~M., {Vandenberg} D.~A., 1998, \aj, 115, 1516

\bibitem[{{Sivarani} {et~al}\mbox{.}(2004){Sivarani}, {Bonifacio}, {Molaro},
  {et~al.}}]{Sivarani2004}
{Sivarani} T., {Bonifacio} P., {Molaro} P., {et~al.}, 2004, \aap, 413, 1073

\bibitem[{{Sneden}(1973)}]{Sneden1973}
{Sneden} C.~A., 1973, PhD thesis, The University of Texas at Austin.

\bibitem[{{Stancliffe} \& {Glebbeek}(2008)}]{Stancliffe2008}
{Stancliffe} R.~J., {Glebbeek} E., 2008, \mnras, 389, 1828

\bibitem[{{Steinhauer}(2003)}]{Steinhauer2003}
{Steinhauer} A., 2003, PhD thesis, Indiana University

\bibitem[{{Takeda} {et~al}\mbox{.}(2008){Takeda}, {Han}, {Kang},
  {et~al.}}]{Takeda2008}
{Takeda} Y., {Han} I., {Kang} D.-I., {et~al.}, 2008, JKAS, 41, 83

\bibitem[{{Thompson} {et~al}\mbox{.}(2008){Thompson}, {Ivans}, {Bisterzo},
  {et~al.}}]{Thompson2008}
{Thompson} I.~B., {Ivans} I.~I., {Bisterzo} S., {et~al.}, 2008, \apj, 677, 556

\bibitem[{{van Dokkum}(2001)}]{Dokkum2001}
{van Dokkum} P.~G., 2001, \pasp, 113, 1420

\bibitem[{{Yang} {et~al}\mbox{.}(2013){Yang}, {Sarajedini}, {Deliyannis},
  {et~al.}}]{Yang2013}
{Yang} S.-C., {Sarajedini} A., {Deliyannis} C.~P., {et~al.}, 2013, \apj, 762, 3

\bibitem[{{Zucker} \& {Mazeh}(1994)}]{Zucker1994}
{Zucker} S., {Mazeh} T., 1994, \apj, 420, 806

\end{thebibliography}

\end{document}